\documentclass[aps,prd,twocolumn,showpacs,groupedaddress,nofootinbib,longbibliography]{revtex4-1}

\usepackage{graphicx}
\usepackage{dcolumn}
\usepackage{amsmath}
\usepackage{amssymb}
\usepackage{xcolor}
\usepackage{multirow}
\usepackage{listings}
\usepackage{xcolor}
\usepackage{lineno}
\usepackage[colorlinks,linkcolor=blue,anchorcolor=blue,citecolor=blue]{hyperref}
\allowdisplaybreaks
\newcommand\dx{{\rm d}}
\newcommand\p{\partial}
\newcommand\etal{{\it et~al.}}

\begin{document}

\title{4G: Pure fourth-order gravity}
\author{Shuxun Tian}
\email[]{tshuxun@whu.edu.cn}
\affiliation{School of Physics and Technology, Wuhan University, 430072, Wuhan, China}
\date{\today}
\begin{abstract}
  Einstein field equations are second-order differential equations. In this paper, we propose a new gravity theory with pure fourth-order field equations, which we call 4G for brevity. We discuss the applications of 4G in cosmology, gravitational waves, and local gravitational systems. 4G predicts the scale factor $a\propto t^{4/3}$ for the matter-dominated universe, and $a\propto t$ for the radiation-dominated universe. The former can explain the late-time acceleration without suffering from the coincidence and fine-tuning problems, while the latter can solve the horizon problem. 4G is a massless gravity, which means the speed of gravitational waves equals to the speed of light. Based on the discussions about exact vacuum solution and weak field approximation, we argue that Schwarzschild metric should be the real physical metric to describe solar system in 4G.
\end{abstract}
\maketitle

\section{Introduction}
The accelerating expansion of the late-time universe is one of the most mysterious phenomena in modern physics. Current standard cosmological model is $\Lambda$-cold-dark-matter ($\Lambda$CDM) model, in which the cosmological constant $\Lambda$ causes the late-time acceleration \cite{Peebles2003,Bartelmann2010}. $\Lambda$ may originate from quantum vacuum energy. So far, $\Lambda$CDM model can explain most mainstream observations \cite{Eisenstein2005,Betoule2014,Ade2016}. However, serious problems exist in the theory. The coincidence and fine-tuning problems are two representatives. The coincidence problem states why the energy density of matter and dark energy is on the same order of magnitude at today \cite{Amendola2010}. The fine-tuning problem arises from the huge difference between the observed dark energy density and the theoretical quantum vacuum energy density \cite{Amendola2010}.

Besides the cosmological constant, current mainstream theories for explaining the late-time acceleration include dark energy models and modified gravity theories (see \cite{DeFelice2010,Sotiriou2010,Capozziello2011,Clifton2012,Joyce2015,Joyce2016} for reviews). Dark energy models add new substances (e.g. quintessence \cite{Caldwell1998,Carroll1998}, k-essence \cite{Armendariz2000} and phantom \cite{Caldwell2002} fields) to the universe within the framework of general relativity\footnote{In this paper, general relativity refers to the original metric gravity with Einstein equations $G_{\mu\nu}=\kappa T_{\mu\nu}$.}. Modified gravity theories (e.g. $f(R)$ gravity \cite{DeFelice2010,Sotiriou2010} and scalar-tensor theory \cite{Brans1961,Wagoner1970}) modify the geometric term in Einstein field equations. To the best of our knowledge, all of these theories retain the Einstein-Hilbert term in the Lagrangian, which is beneficial for them to recover the success of general relativity in the local gravitational systems, e.g. solar system. Modified gravity could be a possible solution to the fine-tuning problem \cite{Bull2016} because it no longer requires quantum vacuum energy. However, all these theories (dark energy models and modified gravity theories) need to introduce parameters in the Lagrangian that relate to the Hubble constant $H_0$. This makes the theory imperfect as the value of $H_0$ depends on the time that humans appear in the universe. In addition, none of these theories can naturally solve the coincidence problem.

What kind of model does not suffer from the coincidence and fine-tuning problems? Here we introduce several non-mainstream models. One class of such models does not modify the expansion dynamics of the universe, and consider the late-time acceleration is just an apparent phenomenon.  Models with dimming of light \cite{Aguirre1999a,Aguirre1999b,Csaki2002} or modified redshift relation \cite{Bassett2013,Wojtak2016,Wojtak2017,Tian2017} belong to this type. However, the former would be excluded by the angular diameter distance measurements \cite{Song2006}, and the latter is not practical because massive existing astronomical data needs to be reprocessed in their own framework as the authors mentioned. Some other models phenomenologically modify the dynamics of the background universe, e.g. power law cosmology \cite{Dolgov2014,Shafer2015}, or more specific $R_{\rm h}=ct$ universe \cite{Melia2012,Melia2018}. From the value of $\chi^2_{\rm min}/{\rm d.o.f.}$, these models perform well in fitting the data. But the problem is these models cannot be used to analyze local gravitational systems due to lack of a complete gravity theory. The most promising theory should be the backreaction theory \cite{Buchert2012}, which retains general relativity and advocates the late-time acceleration is just an effect of cosmological inhomogeneity. The bad news is that the development of backreaction theory is far from mature, and the debate about whether it can work successfully is widely exist in the literatures \cite{Green2014,Buchert2015,Bentivegna2016,Giblin2016,Racz2017,Macpherson2018}.

It is important to find a complete gravity theory, which can explain the late-time acceleration without the coincidence and fine-tuning problems, and recover the successes of general relativity in solar system. In this paper, we try a pure fourth-order gravity approach. The corresponding field equations are pure fourth-order differential equations, which we call 4G for brevity.

This paper is structured as follows. Section \ref{sec:02} explains why 4G is a good theory through the applications on cosmology, gravitational waves, and solar system. Section \ref{sec:03} analyzes the Newtonian approximation, and proposes a physical scenario that enables 4G to recover Newton's law of universal gravitation. This can also be considered as a supplementary material for the applications of 4G in solar system. In Sec. \ref{sec:04}, we propose the final version of 4G theory to solve the problems appear in the previous two sections. Conclusions and discussions can be found in Sec. \ref{sec:05}.

Conventions: $G$ is the gravitational constant, $c$ is the speed of light, $\kappa\equiv8\pi G/c^4$, the metric signature $(-,+,+,+)$. $\p_\mu$ and $\nabla_\mu$ represent partial and covariant derivatives, respectively. $\nabla^\mu\equiv g^{\mu\nu}\nabla_\nu$ and $\Box\equiv\nabla^\mu\nabla_\mu$. The Greek indices run from 0 to 3, and the Latin indices run from 1 to 3. Christoffel symbol $\Gamma^ {\lambda}_{\mu\nu}\equiv g^{\lambda\alpha}(\p_\nu g_{\mu\alpha}+\p_\mu g_{\nu\alpha}-\p_\alpha g_{\mu\nu})/2$, Riemann tensor $R^{\rho}_{\phantom{\rho}{\lambda\mu\nu}}\equiv\p_\mu\Gamma^\rho_{\lambda\nu}- \p_\nu\Gamma^\rho_{\lambda\mu}+\Gamma^\rho_{\alpha\mu}\Gamma^\alpha_{\lambda\nu}-\Gamma^\rho_{\alpha\nu}\Gamma^\alpha_{\lambda\mu}$, Ricci tensor $R_{\mu\nu}\equiv R^\alpha_{\phantom{\alpha}{\mu\alpha\nu}}$, Ricci scalar $R\equiv g^{\mu\nu}R_{\mu\nu}$, and Einstein tensor $G_{\mu\nu}\equiv R_{\mu\nu}-Rg_{\mu\nu}/2$.

\section{Why 4G is good?}\label{sec:02}
What kind of gravity theory is good? In this paper, we emphasize three criteria. Firstly, the theory can explain the late-time acceleration without suffering from the coincidence and fine-tuning problems. Secondly, the theory can explain the dynamics of solar system. More specifically, we expect solar system is described by Schwarzschild metric. Finally, we hope the gravity is massless, which means the speed of gravitational waves equals to $c$. In addition to these three criteria, we also admit two points that Einstein considered when constructing general relativity: Gravity should be a metric theory to explain the weak equivalence principle; energy and momentum conservation can be derived from the field equations.

An action corresponds to a gravity theory. The general form of the action is
\begin{equation}
  S=\int\dx^4x\sqrt{-g}\mathcal{L}_G+S_{\rm m}.
\end{equation}
Note that $\delta S_{\rm m}=\int\dx^4x\sqrt{-g}T_{\mu\nu}\delta g^{\mu\nu}/2$ \cite{Gasperini2013}. In this paper, we only consider the perfect fluid, whose energy-momentum tensor $T_{\mu\nu}=(\rho+p/c^2)u_\mu u_\nu+pg_{\mu\nu}$, where $\rho$ is density, $p$ is pressure. The variational method ensures energy conservation can be derived from the field equations. We completely abandon Einstein-Hilbert term in the Lagrangian, because it leaves a second-order differential term in the field equations. In order to obtain pure fourth-order field equations, the simplest Lagrangian is
\begin{equation}
  \mathcal{L}_G=\frac{1}{4\zeta}R^2,
\end{equation}
where $\zeta$ is the coupling constant with dimension of ${\rm s}^2\cdot{\rm kg}^{-1}\cdot{\rm m}^{-3}$. Palatini formalism is not in our consideration because the corresponding field equations are not suitable for describing low energy gravity \cite{Sotiriou2010}. Using metric formalism in the $f(R)$ theories, we obtain the field equations (see \cite{DeFelice2010,Sotiriou2010} or Eq. (286) and Eq. (288) in \cite{Clifton2012})
\begin{equation}\label{eq:03}
  F_{\mu\nu}\equiv RR_{\mu\nu}-\frac{g_{\mu\nu}}{4}R^2-\left[\nabla_\mu\nabla_\nu-g_{\mu\nu}\Box\right]R=\zeta T_{\mu\nu}.
\end{equation}
However, Eq. (\ref{eq:03}) also cannot be used to describe low energy gravitational systems, e.g. solar system. The reason is here. On the one hand, contraction of Eq. (\ref{eq:03}) gives $3\Box R=\zeta T$. On the other hand, for the static weak gravitational systems, the only non-zero energy-momentum tensor is $T_{00}=\rho c^4$ (see Sec. \ref{sec:0301} for details) and there should be a time-independent solution for the field equations. The first two terms of Eq. (\ref{eq:03}) contribute nothing to the linear approximation as the background is Minkowski metric. Then 00 component gives $-c^2\Box R=\zeta\rho c^4$, i.e. $\Box R=T$. This is contrary to the general result. Therefore, Eq. (\ref{eq:03}) does not represent a good gravity theory.

The flaw of Eq. (\ref{eq:03}) is that only $R$, not $4\times4$ geometric components, appears in the linearized field equations. We can choose other Lagrangian to solve this problem. One simple choice is
\begin{equation}\label{eq:04}
  \mathcal{L}_G=\frac{1}{2\zeta}R_{\mu\nu}R^{\mu\nu}.
\end{equation}
Variation of the action with respect to $g_{\mu\nu}$ gives the field equations (see \cite{Matyjasek2011} or Eqs. (368--372) in \cite{Clifton2012})
\begin{align}\label{eq:05}
  H_{\mu\nu}\equiv\ &\Box R_{\mu\nu}+\frac{g_{\mu\nu}}{2}\Box R-\nabla_\mu\nabla_\nu R\nonumber\\
  &-2R_{\alpha\mu\nu\beta}R^{\alpha\beta}-\frac{g_{\mu\nu}}{2}R_{\alpha\beta}R^{\alpha\beta}=\zeta T_{\mu\nu}.
\end{align}
Energy and momentum conservation in curved spacetime can be derived from the above equations, which means test particles move along the geodesic \cite{Gasperini2013}. We only introduce one parameter $\zeta$ into the field equations, which will be determined by Newtonian approximation. Before that, we first discuss the good properties of Eq. (\ref{eq:05}) in cosmology, gravitational waves, and local gravitational systems.

\subsection{Cosmology}
The cosmological principle states the universe is homogeneous and isotropic. In addition, we assume the universe is flat. Then the universe could be described by the flat Friedmann-Lema\^{i}tre-Robertson-Walker (FLRW) metric
\begin{equation}\label{eq:06}
  \dx s^2=-c^2\dx t^2+a^2\dx\mathbf{r}^2,
\end{equation}
where $\dx\mathbf{r}^2=\dx x^2+\dx y^2+\dx z^2$, and $a=a(t)$ is the scale factor of the universe. The corresponding energy-momentum tensor $T_{00}=\rho c^4$, $T_{0i}=0$ and $T_{ij}=\delta_{ij}pa^2$. We search for the power law cosmology with
\begin{equation}\label{eq:07}
  a=a_0\left(\frac{t}{t_0}\right)^n.
\end{equation}
Substituting Metric (\ref{eq:06}) into Eq. (\ref{eq:05}), 00 and 11 components give \cite{MapleCode}
\begin{gather}
  -\frac{18n^2(2n-1)}{c^2t^4}=\zeta\rho c^4,\label{eq:08}\\
  \frac{6a^2n(6n^2-11n+4)}{c^4t^4}=\zeta pa^2,\label{eq:09}
\end{gather}
respectively. In principle, we can directly determine the value of $n$ according to energy conservation and Eq. (\ref{eq:08}). But in order to verify the correctness of the {\tt Maple} Program \cite{MapleCode}, we adopt the calculation completely based on the gravitational field equations.

Solving Eq. (\ref{eq:09}) gives $n=4/3$ for the pressureless matter, which means the expansion of the  matter-dominated universe is accelerating. This solution corresponding to $\rho\propto a^{-3}$, i.e. energy conservation. Comparing power law cosmology with $\Lambda$CDM model, observations may favor the latter one \cite{Shafer2015}. However, the power law cosmology with $a\propto t^{4/3}$ does fit the data well if we only look at the value of $\chi^2_{\rm min}/{\rm d.o.f.}$ (see the constraint results in \cite{Dolgov2014,Shafer2015}). In this paper, we tacitly approve $a\propto t^{4/3}$ can successfully explain the observed late-time acceleration, and do not care more details about cosmological parameter constraints. From a theoretical point of view, $a\propto t^{4/3}$ can naturally solve the coincidence problem. No parameter related to $H_0$ introduced in Eq. (\ref{eq:05}) and the fine-tuning problem disappears in this theory. Combined with $n=4/3$, Eq. (\ref{eq:08}) gives $\zeta<0$.

For radiation, $p=\rho c^2/3$ and the solution of Eq. (\ref{eq:09}) is $n=1$, which corresponds to $\rho\propto a^{-4}$ as we expected. We can use $a\propto t$ in the radiation-dominated era to solve the horizon problem\footnote{Horizon problem arises from the classical Big Bang cosmology cannot explain the homogeneity of the cosmic microwave background temperature in the causally disconnected space regions. If $a\propto t$ for the radiation-dominated era, then the whole region of the last scattering surface observed at today is causally connected at early time.}. However, the flatness problem is still puzzling if we completely abandon the inflation period \cite{Guth1981}.

\subsection{Gravitational waves}
In general, adding $R^2$ or $R_{\mu\nu}R^{\mu\nu}$ terms to the Einstein-Hilbert action always makes the gravity to be massive \cite{Clifton2012}. For example, in the $f(R)=R+\alpha R^2$ theory, contraction of the vacuum field equations gives $(\Box-m_0^2)R=0$, where $m_0^2=1/(6\alpha)$. This result indicates the existence of a massive gravitational wave component \cite{Liang2017}. However, things changed if we only retain $R_{\mu\nu}R^{\mu\nu}$ term in the Lagrangian. Using the antisymmetric property of Riemann tensor $R_{\alpha\mu\nu\beta}=-R_{\mu\alpha\nu\beta}$, we obtain  $g^{\mu\nu}R_{\alpha\mu\nu\beta}R^{\alpha\beta}=-R_{\alpha\beta}R^{\alpha\beta}$. So in vacuum, contraction of Eq. (\ref{eq:05}) gives
\begin{equation}
  \Box R=0,
\end{equation}
which means this gravity is massless and the speed of gravitational waves equals to $c$. A single gravitational wave event could give tight bound on the graviton mass through dispersion observations \cite{GW150914}. The joint observations of gravitational waves and their electromagnetic counterparts can give a strong limit on the absolute speed of gravitational waves \cite{GW170817}. From a theoretical point of view, we can fully believe gravity is massless and the speed of gravitational waves equals to $c$, otherwise a small parameter is needed, which may lead to a new fine-tuning problem.

\subsection{Solar system: Schwarzschild metric}\label{sec:0203}
General relativity has achieved great success in solar system \cite{Will2014}. But be aware that most of current tests can only be regarded as a verification of Schwarzschild metric. Thus, for the first step, it is crucial to judge whether Schwarzschild metric is a solution for Eq. (\ref{eq:05}). Fortunately, $R_{\mu\nu}=0$ represents a special kind of solution for Eq. (\ref{eq:05}), i.e. Schwarzschild metric is still an exact vacuum solution. However, it is not enough that just proving Schwarzschild metric is an exact solution. The problem is in many massive gravity theories, solar system is not described by Schwarzschild metric, even though it is an exact vacuum solution. What we need is not only to prove Schwarzschild metric is an exact solution, but also to prove the weak field approximation solution is consistent with Schwarzschild metric\footnote{In general relativity, Birkhoff theorem also guarantees solar system should be described by Schwarzschild metric. But for a general gravity theory, one must analyze these two issues.}. Here we use $f(R)=R+\alpha R^2$ theory to clarify this statement. In this theory, the Newtonian gravitational potential for a point mass can be written as (see Eq. (307) in \cite{Clifton2012})
\begin{equation}
  \Phi_{f(R)}=\frac{c_1}{r}\cdot\left(1+\frac{1}{3}e^{-m_0r}\right).
\end{equation}
where $c_1$ is constant. In contrast, the Newtonian gravitational potential in 4G is (see Sec. \ref{sec:0302})
\begin{equation}
  \Phi_{\rm 4G}=\frac{c_1}{r}.
\end{equation}
Linearizing Schwarzschild metric, we obtain the corresponding Newtonian gravitational potential $\Phi_{\rm Sch}=-GM/r$, where $M$ is the total mass. Note that Newtonian potential is related to the perturbation of $g_{00}$ (see Sec. \ref{sec:0301}). It is clear to see that $\Phi_{f(R)}$ is not consistent with the linearized Schwarzschild metric. Therefore, in $f(R)=R+\alpha R^2$ theory, solar system should not be described by Schwarzschild metric. However, $\Phi_{4G}$ is consistent with the linearized Schwarzschild metric. Taking into account these discussions, we believe solar system should be described by Schwarzschild metric in the framework of 4G.

\section{Newtonian approximation}\label{sec:03}
Whether or not a gravity theory can be accepted depends largely on the performance of its weak field approximation. The analysis of weak field approximation not only helps to pick out the right vacuum solution for a real gravitational system as we discussed in the last section, but also helps to determine the value of the coupling constant in the field equations. In this section, we first review the Newtonian approximation in general relativity and then analyze the case of 4G.

\subsection{General relativity}\label{sec:0301}
Newtonian approximation of general relativity has been widely exist in the literatures \cite{Gasperini2013,Misner1973}. We just restate the main results. The motivation for doing this is mainly to find out which hypothesis can be made in the Newtonian approximation analysis, so as not to make a wrong assumption in 4G. We write the metric as $g_{\mu\nu}=\bar{g}_{\mu\nu}+h_{\mu\nu}$. The bar represents the background metric and its related quantity; $h_{\mu\nu}$ is the perturbation and satisfies $h_{\mu\nu}=h_{\nu\mu}$. Based on the definition, we obtain
\begin{align}
  &\delta R_{\mu\nu}\equiv R_{\mu\nu}(g)-R_{\mu\nu}(\bar{g})\nonumber\\
  &\ =\frac{1}{2}\left(-\bar{\Box}h_{\mu\nu}
  -\bar{\nabla}_\mu\bar{\nabla}_\nu h^\alpha_{\phantom{\alpha}{\alpha}}
  +\bar{\nabla}_\mu\bar{\nabla}_\alpha h^\alpha_{\phantom{\alpha}{\nu}}
  +\bar{\nabla}_\nu\bar{\nabla}_\alpha h^\alpha_{\phantom{\alpha}{\mu}}\right.\nonumber\\
  &\qquad\quad\left.-2\bar{R}^{\alpha\phantom{\mu}{\beta}}_{\phantom{\alpha}{\mu\phantom{\beta}{\nu}}}h_{\alpha\beta}
  +\bar{R}^\alpha_{\phantom{\alpha}{\mu}}h_{\alpha\nu}+\bar{R}^\alpha_{\phantom{\alpha}{\nu}}h_{\mu\alpha}\right),\label{eq:13}\\
  &\delta R\equiv R(g)-R(\bar{g})\nonumber\\
  &\ =-\bar{\Box}h^\alpha_{\phantom{\alpha}{\alpha}}+\bar{\nabla}^\alpha\bar{\nabla}_\beta h^\beta_{\phantom{\beta}{\alpha}}-\bar{R}^{\alpha\beta}h_{\alpha\beta},\label{eq:14}
\end{align}
where $\bar{\nabla}_\alpha$ is the covariant derivative defined on $\bar{g}_{\mu\nu}$, $\bar{\nabla}^\alpha\equiv\bar{g}^{\alpha\beta}\bar{\nabla}_\beta$, $\bar{\Box}\equiv\bar{\nabla}^\alpha\bar{\nabla}_\alpha$, and $h^\mu_{\phantom{\mu}{\nu}}\equiv\bar{g}^{\mu\alpha}h_{\alpha\nu}$. So far, we have not set any limits on the background metric. In the Newtonian approximation analysis, we can choose Minkowski metric as the background.

Here we first analyze the motion of test particles. In the Minkowski background, combined with the weak field and low speed approximation, geodesic equation gives \cite{Misner1973}
\begin{equation}\label{eq:15}
  \frac{\dx^2x^i}{\dx t^2}\approx\frac{1}{2}\frac{\p h_{00}}{\p x^i}.
\end{equation}
A key point in the derivation is $\p_0h_{0i}\ll\p_i h_{00}$, which would limit the math form of the gauge. Note that Eq. (\ref{eq:15}) is only valid in Cartesian coordinates. Combined with Newtonian mechanics, we obtain the Newtonian potential $\Phi(\vec{x})=-h_{00}/2+{\rm const}$. Without loss of generality, we can set this constant to zero.

Now we back to Einstein field equations. The first step is to simplify the energy-momentum tensor. The source is assumed to be static. Combined with the weak field assumption (density and pressure are small), we can adopt the four-velocity $u^\mu=(1,0,0,0)$. In order to keep the source static, pressure should exist to resist gravity. However, pressure term is a higher-order infinitesimal because $p/c^2\ll\rho$. This is consistent with the fact that pressure does not contribute to Newtonian gravity. So at linear level, the only non-zero energy-momentum tensor is $T_{00}=\rho c^4$. $\dx\rho/\dx t=0$ represents energy conservation. Next, we search for the relation between $h_{00}$ and $\rho$. Because $\nabla^\nu G_{\mu\nu}=0$, only six of ten field equations are independent of each other. This allows us to add four restrictions: $\bar{\nabla}_\nu h^\nu_{\ \mu}=0$, which greatly simplify the expressions of $\delta R_{\mu\nu}$ and $ \delta R$. However, $G_{00}=\kappa T_{00}$ still cannot give what we want, because $h^\alpha_{\ \alpha}\neq0$. Note that we are not allowed to set $h^\alpha_{\ \alpha}\neq0$, otherwise one would obtain $\rho\propto T\propto R=0$ based on Eq. (\ref{eq:14}). In order to eliminate $h^\alpha_{\ \alpha}$, we should contract the field equations. Equivalently, we can start from Einstein equations in the form of $R_{\mu\nu}=\kappa(T_{\mu\nu}-Tg_{\mu\nu}/2)$. Because the source is static, we can assume there exist a time-independent solution for $h_{\mu\nu}$. Combined with Eq. (\ref{eq:13}), the 00 component of the field equations gives
\begin{equation}
  -\Delta h_{00}=\kappa\rho c^4,
\end{equation}
where $\Delta$ is the Laplace operator. Combined with Poisson equation $\Delta\Phi=4\pi G\rho$ and $\Phi=-h_{00}/2$, we obtain $\kappa=8\pi G/c^4$ in general relativity.

The above analysis gives the equation that $ h_ {00} $ satisfies without any information about $ h_ {i\mu} $. There is a more straightforward method of computation: Using an explicit perturbation form, i.e. a specific gauge, to directly simplify the gravitational field equations. A widely used gauge is the Newtonian gauge with a constant scale factor in the FLRW metric \cite{Clifton2012}
\begin{equation}\label{eq:17}
  \dx s^2=-c^2(1+2\Phi/c^2)\dx t^2+(1-2\Psi/c^2)\dx\mathbf{r}^2,
\end{equation}
where $\Phi=\Phi(\vec{x})$ and $\Psi=\Psi(\vec{x})$. If we set $\Psi=\Phi$, then $G_{i\mu}=0=\kappa T_{i\mu}$, and the 00 component gives the Poisson equation. If we further set $\Phi=-GM/r$, then Gauge (\ref{eq:17}) is equivalent to the linearized Schwarzschild metric. In general, $\Psi\neq\Phi$ for the massive gravity theories (see examples illustrated in \cite{Clifton2012}). Therefore, in these theories, Gauge (\ref{eq:17}) cannot be consistent with the linearized Schwarzschild metric. This could be regarded as a supplementary solution to the issue we discussed in Sec. \ref{sec:0203}: In a massive gravity, why solar system is not described by Schwarzschild metric, while it is an exact vacuum solution?

Another gauge worth mentioning is the synchronous-like gauge\footnote{Synchronous gauge has been widely used in cosmological perturbation analysis \cite{Ma1995}, in which $h_{0\mu}=0$. Here we only discuss the special case with $h_{00}=0$.}, in which $h_{00}=0$. What is wrong with Eq. (\ref{eq:15}) if $h_{00}=0$? First of all, we declare there is no contradiction in the theory. To explain this, we start from the linearized Schwarzschild metric $\dx s^2=-c^2(1-r_s/r)\dx t^2+(1+r_s/r)\dx r^2+r^2\dx\Omega^2$, where $r_s$ is the Schwarzschild radius. Taking coordinate transformation about time $t\rightarrow\tilde{t}=[1-r_s/(2r)]t$, we obtain $\dx s^2=-c^2\dx t^2+c^2tr_s /r^2\dx t\dx r+(1+r_s/r)\dx r^2+r^2\dx\Omega^2$, in which we rewrite $\tilde{t}$ as $t$. Transforming spherical coordinates into Cartesian coordinates does not change anything about what we cares. In this form $h_{00}=0$. But $h_{0i}$ depends on time, which is contrary to the assumption we made when deriving Eq. (\ref{eq:15}). Previous contradiction stems from we apply incompatible assumptions into one analysis. For a static gravity source, it is naturally to believe that there exist a time-independent solution for the field equations. Eq. (\ref{eq:15}) indicates once $h_{\mu\nu}$ are independent of time is assumed, we can no longer assume $h_{00}=0$.

\subsection{$R_{\mu\nu}R^{\mu\nu}$ gravity}\label{sec:0302}
Now we study the static weak gravitational field systems in 4G. For the source, the only non-zero energy-momentum tensor is $T_{00}=\rho c^4$. Minkowski metric is still an exact vacuum solution for Eq. (\ref{eq:05}), and we can choose it as background. In the last subsection, we introduced two methods to linearize Einstein field equations: One is finding a general expression of $\delta R_{\mu\nu}$, and the other one is directly calculating Einstein tensors with Gauge (\ref{eq:17}). Here we take the calculation more straightforward method, i.e. simplify Eq. (\ref{eq:05}) with a specific gauge. But one thing should keep in mind is that the choice of gauge may depends on the gravity theory. Gauge (\ref{eq:17}) with $\Psi=\Phi$ may not work for Eq. (\ref{eq:05}). The desired gauge should satisfy the following four criteria:
\begin{enumerate}
  \item Metric perturbation is independent of time;
  \item $h_{00}=-2\Phi$;
  \item $H_{i\mu}=0$ is automatically established at linear level;
  \item In areas far from the source, the gauge is consistent with the linearized Schwarzschild metric.
\end{enumerate}
Here we would like to discuss more about the fourth criterion. Schwarzschild metric is obtained from solving Einstein equations. This metric has been widely tested through various observations \cite{Will2014}, or at the linear level, observations show the ratio $\Psi/\Phi$ is very close to 1 \cite{Bertotti2003,Collett2018}. In our opinion, Einstein field equations can be completely abandoned, but Schwarzschild metric should be retained. As we discussed in Sec. \ref{sec:0203},  in order for solar system to be described by Schwarzschild metric, two conditions are necessary: Schwarzschild metric is an exact vacuum solution; weak field approximation solution is equivalent to the linearized Schwarzschild metric. This is why we demand the fourth criterion.

Substituting Gauge (\ref{eq:17}) into Eq. (\ref{eq:05}), we find $H_{i\mu}=\widehat{\rm O}_{i\mu}(\Phi-3\Psi)$ \cite{MapleCode}, where $\widehat{\rm O}_{i\mu}$ is differential operator. Therefore, setting $\Psi=\Phi/3$ will satisfy the third criterion. However, this result violates the fourth criterion. Fortunately, we find that adding $1/r$ to the diagonal does not affect $H_{\mu\nu}$ \cite{MapleCode}. The linearized geodesic equation does not allow us to add extra $1/r$ term to $g_{00}$. Therefore, in order to meet the fourth criterion, we can write the gauge as
\begin{equation}\label{eq:18}
  \dx s^2=-c^2(1+\frac{2\Phi}{c^2})\dx t^2+(1-\frac{2\Phi}{3c^2}+\frac{4GM}{3c^2r})\dx\mathbf{r}^2,
\end{equation}
where $M$ is the total mass of the local gravitational system. However, Gauge (\ref{eq:18}) is not a good gauge due to the divergence of $1/r$ at $r=0$. We will solve this problem in the next section. As a preview, we construct new field equations and Gauge (\ref{eq:17}) with $\Psi=\Phi$ is valid in the corresponding Newtonian approximation analysis. For now, Gauge (\ref{eq:18}) is enough for us to discuss a lot of things.

Substituting Gauge (\ref{eq:18}) into Eq. (\ref{eq:05}), 00 component gives the linearized gravitational field equation \cite{MapleCode}
\begin{equation}\label{eq:19}
  \frac{4}{3}\Delta^2\Phi=\zeta\rho c^4,
\end{equation}
where $\Delta^2=\Delta\Delta$. For a spherically symmetric system, the above equation can be simplified to
\begin{equation}\label{eq:20}
  \frac{\dx^4\Phi}{\dx r^4}+\frac{4}{r}\frac{\dx^3\Phi}{\dx r^3}=\frac{3\zeta c^4}{4}\rho.
\end{equation}
The vacuum solution is
\begin{equation}\label{eq:21}
  \Phi(r)=\frac{c_1}{r}+c_2+c_3r+c_4r^2,
\end{equation}
where $c_i$ is constant. The first term could be used to recover Newton's gravity. The second term is a constant and can be neglected. The real physical solution should vanish at infinity. This boundary condition at infinity has been used to suppress the $\exp(+m_0r)$ term in the $f(R)=R+\alpha R^2$ theory \cite{Pechlaner1966}. For Eq. (\ref{eq:21}), we have $c_3=0$ and $c_4=0$. Now, we focus on constructing a physical scenario makes Eq. (\ref{eq:19}) and Newton's gravity equivalent. Obviously, Eq. (\ref{eq:19}) is not equivalent to Poisson equation. However, as we will see, Poisson equation is not necessary for recovering Newtonian gravity. The $1/r$ term appears in Eq. (\ref{eq:21}) indicates recovering Newton's gravity from Eq. (\ref{eq:19}) is possible.

Before dealing with Eq. (\ref{eq:19}), here we discuss how to determine the coupling constant in Poisson equation. The Poisson equation can be written as $\Delta\Phi=A\rho$, where $A$ is the coupling constant. To determine $A$, a rigorous way is to compare the Green function solution with the desired expression of Newtonian potential for a continuously distributed source. A simpler method is to solve Poisson equation for a specific matter density distribution, and then tuning $A$ to obtain the desired Newtonian potential for distant areas. For example, we can assume the density satisfies a spherically symmetric three-dimensional normal distribution
\begin{equation}\label{eq:22}
  \rho(r)=\frac{M}{(2\pi)^{3/2}\sigma^3}\exp(-\frac{r^2}{2\sigma^2}),
\end{equation}
where $M$ is the total mass and $\sigma$ is a positive constant in dimension of meter. We expect $\Phi$ approach to $-GM/r$ as $r$ goes to infinity. Solving Poisson equation gives
\begin{equation}
  \Phi=-\frac{AM}{4\pi r}{\rm erf}(\frac{r}{\sqrt{2}\sigma})+\frac{c_1}{r}+c_2,
\end{equation}
where ${\rm erf}(x)$ is the error function. When $r\rightarrow0$, the first term just contributes one constant due to ${\rm erf}(x)=2/\sqrt{\pi}(x-x^3/3+x^5/10-\cdots)$. In order for $\Phi(r)$ to be continuous at $r=0$, we should set $c_1=0$. The constant term can be neglected. We know $\lim_{x\rightarrow+\infty}{\rm erf}(x)=1$. Then the above solution gives $\Phi=-AM/(4\pi r)$ for large $r$. Compared to the desired Newtonian potential $\Phi=-GM/r$, we know $A=4\pi G$, which is exactly the result given by Green function method.

Now we apply the above method to Eq. (\ref{eq:19}). The general solution for $\Delta^2\Phi=A\rho$ with Eq. (\ref{eq:22}) is
\begin{align}\label{eq:24}
  \Phi=&-\frac{AMr}{8\pi}{\rm erf}(\frac{r}{\sqrt{2}\sigma})
  -\frac{AM\sigma^2}{8\pi r}{\rm erf}(\frac{r}{\sqrt{2}\sigma})+\frac{c_1}{r}\nonumber\\
  &-\frac{\sqrt{2}AM\sigma}{8\pi^{3/2}}\exp(-\frac{r^2}{2\sigma^2})+c_2+c_3r+c_4r^2.
\end{align}
The boundary condition at infinity gives $c_3=AM/(8\pi)$ and $c_4=0$. When $r$ is large, the $r$ term could be completely abandoned due to $[1-{\rm erf}(x)]x\ll1/x$. In order for $\Phi(r)$ to be continuous at $r=0$, we should also set $c_1=0$. The constant term can be neglected. Thus, in the areas away from the center, the above solution can be simplified to
\begin{equation}\label{eq:25}
  \Phi=-\frac{AM\sigma^2}{8\pi r}.
\end{equation}
Compared to the desired Newtonian potential, we obtain $A\sigma^2/(8\pi)=G$, i.e. $\zeta=32\pi G/(3c^4\sigma^2)$. Note that $A=3\zeta c^4/4$ in Eq. (\ref{eq:25}). This result indicates $\zeta>0$, which is contrary to the result given by Eq. (\ref{eq:08}). We will solve this problem in the next section.

$\zeta$ can be regarded as a fundamental physical constant that represents the coupling strength between matter and spacetime. $\sigma$ is just a parameter we introduced artificially in Eq. (\ref{eq:22}). Then one problem arises, what is wrong with $\zeta=32\pi G/(3c^4\sigma^2)$? Before answering this question, we return to Eq. (\ref{eq:25}), which is proportional to $\sigma^2$. This result seems to mean that Newtonian gravity depends not only on mass but also on density dispersion (or size). However, two facts make us believe that this conclusion does not hold for macroscopic objects. One is a theoretical consideration: The macro interpretation of $\Phi\propto\sigma^2$ is contrary to the microscopic interpretation to be introduced in the next paragraph. Microscopic physical mechanism should be more basic, and macroscopic law should not be contrary to the microscopic mechanism. The other fact comes from the experiment. If $\Phi\propto\sigma^2$ is valid for macro objects, then the value of $G$ measured by ground-based experiments should depend on the elemental composition, density, and size etc. However, current experiments does not find this dependency \cite{Cavendish1798,Luther1982,Gundlach2000,Armstrong2003,Luo2009,Tu2010,Rothleitner2017}. Therefore, we conclude $\Phi\propto\sigma^2$ does not apply to macro objects.

\begin{figure*}
	\includegraphics[width=1.98\columnwidth]{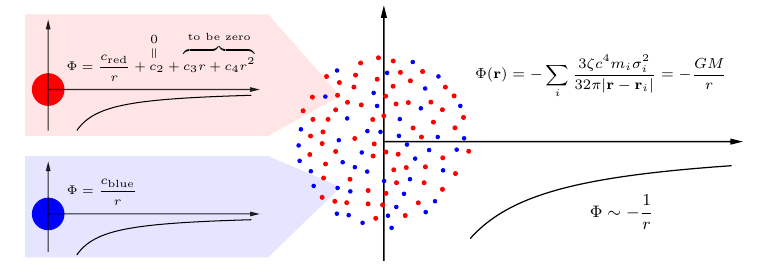}
    \caption{Illustration of a microscopic mechanism that recover Newton's law of universal gravitation from 4G. Different colors represent different fundamental particles, whose mass $m$ and density variance $\sigma^2$ can be different. The meaning of each symbol can be found in the main text. For simplicity, we assume the distribution of particles is spherically symmetric. The gravitational potential generated by each particle is calculated by Eq. (\ref{eq:25}). Note that $r$ and $r^2$ terms are set to be zero according to the boundary condition at infinity. The total gravitational potential can be obtained by summing over all the particles, because Eq. (\ref{eq:19}) is a linear differential equation.}
    \label{fig:01}
\end{figure*}

What if $\Phi\propto\sigma^2$, i.e. Eq. (\ref{eq:25}), is only valid for microscopic particles? Keeping this idea in mind, here we propose a physics scenario that recovers Newton's law of universal gravitation from Eq. (\ref{eq:19}) and Eq. (\ref{eq:25}). As illustrated in Fig. \ref{fig:01}, matter is composed of fundamental particles. Different particles may own different $m$ and $\sigma^2$. The Newtonian potential for each particle is calculated by Eq. (\ref{eq:25}). For simplicity, we assume particles are distributed spherically symmetric in space and localized in a certain size. Eq. (\ref{eq:19}) is a linear differential equation, which allows us to add Newtonian potential of each particle to obtain the total gravitational potential
\begin{equation}
  \Phi(\mathbf{r})=-\sum_i\frac{3\zeta c^4m_i\sigma_i^2}{32\pi|\mathbf{r}-\mathbf{r}_i|}=-\frac{GM}{r},
\end{equation}
where $i$ sum over all particles, the second equality is only valid outside, $M=\sum_im_i$, and
\begin{equation}\label{eq:27}
  G=\frac{3\zeta c^4\langle\sigma_i^2\rangle}{32\pi}=\frac{3\zeta c^4}{32\pi M}\sum_im_i\sigma^2_i.
\end{equation}
The above formula means $G$ is an average value, not a fundamental physical constant. In addition, Eq. (\ref{eq:27}) depends on the specific mathematical form of Eq. (\ref{eq:22}). However, different forms are just changing a dimensionless constant. If we further assume fundamental particles share same proportion in various macro objects, then the measured value of $G$ should be independent of elemental composition (at atomic level), density, and size etc. Especially, this assumption is reasonable if we consider fundamental particles as electrons, quarks, or some more basic particles. In summary, using the above physical scenario, we can obtain Newton's gravity from Eq. (\ref{eq:19}).

\section{Problems and solutions for Eq. (\ref{eq:05})}\label{sec:04}
Until now, all the discussions about 4G is based on Eq. (\ref{eq:05}). Two problems has been mentioned before. One is the divergence problem of Gauge (\ref{eq:18}) at $r=0$. The other one is Eq. (\ref{eq:08}) and Eq. (\ref{eq:25}) gives opposite sign of $\zeta$. In this section, we construct new gravitational field equations to solve these two problems.

A general Lagrangian for the fourth-order field equations can be written as $\mathcal{L}_G=c_1R^2+c_2R_{\mu\nu}R^{\mu\nu}+c_3R_{\alpha\mu\nu\beta}R^{\alpha\mu\nu\beta}$. We can ignore the last term due to the Gauss-Bonnet theorem. Without loss of generality, we can fix one coefficient. We set $c_2=1/(2\zeta)$. Tuning $c_1$ may help us achieve our goals. Actually and luckily, it does work with $c_1=-1/(4\zeta)$. For summary, the Lagrangian should be
\begin{equation}\label{eq:28}
  \mathcal{L}_G=\frac{1}{2\zeta}R_{\mu\nu}R^{\mu\nu}-\frac{1}{4\zeta}R^2,
\end{equation}
and the corresponding field equations are $H_{\mu\nu}-F_{\mu\nu}=\zeta T_{\mu\nu}$, i.e.
\begin{align}\label{eq:29}
  \Box R_{\mu\nu}-\frac{g_{\mu\nu}}{2}\Box R&
  +\frac{g_{\mu\nu}}{4}R^2-\frac{g_{\mu\nu}}{2}R_{\alpha\beta}R^{\alpha\beta}\nonumber\\
  &-2R_{\alpha\mu\nu\beta}R^{\alpha\beta}-RR_{\mu\nu}=\zeta T_{\mu\nu}.
\end{align}
The first two terms are just $\Box G_{\mu\nu}$ once we remember $\nabla_\alpha g_{\mu\nu}=0$. Here we briefly summarize the good properties of Eq. (\ref{eq:29}) in cosmology, gravitational waves, and local gravitational systems.
\begin{itemize}
  \item Cosmology. Substituting Metric (\ref{eq:06}) and Eq. (\ref{eq:07}) into Eq. (\ref{eq:29}), we obtain \cite{MapleCode}
      \begin{gather}
        \frac{9n^2(2n-1)}{c^2t^4}=\zeta\rho c^4,\label{eq:30}\\
        -\frac{3a^2n(6n^2-11n+4)}{c^4t^4}=\zeta pa^2.\label{eq:31}
      \end{gather}
      This gives $a\propto t^{4/3}$ for the matter-dominated universe, and $a\propto t$ for the radiation-dominated universe. In the view of cosmology, $a\propto t^{4/3}$ can explain the late-time acceleration without suffering from the coincidence and fine-tuning problems. $a\propto t$ can solve the horizon problem exist in the classical Big Bang cosmology. However, $a\propto t$ cannot solve the flatness problem. Eq. (\ref{eq:30}) gives $\zeta>0$.
  \item Gravitational waves. In the vacuum region, contraction of Eq. (\ref{eq:29}) gives $\Box R=0$. This means 4G is massless and the speed of gravitational waves equals to $c$.
  \item Schwarzschild metric. Firstly, Schwarzschild metric is still an exact vacuum solution. Secondly, Gauge (\ref{eq:17}) with $\Psi=\Phi$ satisfies the four conditions listed in Sec. \ref{sec:0302} and is applicable to Eq. (\ref{eq:29}). These two reasons make us believe that solar system should still be described by Schwarzschild metric in 4G (see more discussions in Sec. \ref{sec:0203} and Sec. \ref{sec:03}). Substituting Gauge (\ref{eq:17}) with $\Psi=\Phi$ into Eq. (\ref{eq:29}), we obtain the linearized gravitational field equation \cite{MapleCode}
      \begin{equation}\label{eq:32}
        2\nabla^4\Phi=\zeta\rho c^4.
      \end{equation}
      Executing a calculation similar to Eq. (\ref{eq:19}), we get $\zeta=16\pi G/(c^4\sigma^2)$. A more rigorous expression similar to Eq. (\ref{eq:27}) is $G=\zeta c^4\langle\sigma_i^2\rangle/(16\pi)$. This result shows $\zeta>0$, which is consistent with the cosmological result. Section \ref{sec:0302} describes a physical scenario that allows us to recover Newton's law of gravity from Eq. (\ref{eq:32}).
\end{itemize}

\section{Conclusions and discussions}\label{sec:05}
In this paper, we propose 4G --- a gravity theory with pure fourth-order differential equations. Eq. (\ref{eq:29}) is the core of this paper. In 4G, material-dominated universe can explain the late-time acceleration without suffering from the coincidence and fine-tuning problems. The horizon problem disappears in the radiation+matter universe. 4G predicts the speed of gravitational waves equals to $c$. We present a full consistent perturbation analysis around the Minkowski background and show 4G do can recover Newton's law of universal gravitation in the weak field limit.

It is possible to extend 4G. In the framework of 4G, one can add terms like $\Box R$ \cite{Schmidt1990} to the Lagrangian. For higher-order theories, one can try $\mathcal{L}_G=\Box^2R$ or $R_{\mu\nu}\Box R^{\mu\nu}$ etc. However, $\mathcal{L}_G=R^3$ is forbidden because perturbation of the corresponding field equations in the Minkowski background cannot give linear differential equations. Note that the linearity of Eq. (\ref{eq:32}) is critical to recovering Newtonian gravity.

Here we would like to discuss more about the status of current gravity research. Conservatively speaking, there are hundreds of modified gravity theories. Some of them introduce a scalar field in the Lagrangian, e.g. Brans-Dicke theory \cite{Brans1961}, while some of them just modify the geometry part, e.g. $f(R)$ theory \cite{DeFelice2010,Sotiriou2010}. A new scalar field may be popular in high-energy physics, but it may not be in gravity research (see the comment of Hawking and Ellis \cite{Hawking1973}). For the theory that just modify the geometry part, generally, in addition to a parameter related to $G$, there are other parameters related to $H_0$ in the Lagrangian. Observational constraints can determine the values of these parameters. But a more elegant way should be to determine these parameters based on the theory's own considerations. $f(R)$ cannot do this, while Lagrangian (\ref{eq:28}) can. The key is we demand gravity theory should adopt Gauge (\ref{eq:17}) with $\Psi=\Phi$. This is reasonable, because observations do impose strong limit on the deviation of $\Psi=\Phi$ \cite{Bertotti2003}. From a theoretical point of view, we can trust $\Psi=\Phi$ and use it to pick out the beautiful gravity theory.

\section*{Acknowledgements}
ST acknowledges Miss Carrot for discussions about general relativity and company during the entire work. This work was supported by the National Natural Science Foundation of China under Grants No. 11633001.

%\bibliography{bibfile}

\clearpage
\newpage
%%%%%%%%%%%%%%%%%%%%%%%%%%%% .sty for Maple code %%%%%%%%%%%%%%%%%%%%%%%%%
\definecolor{emphcolor}{rgb}{0.58,0,0.82}
\lstdefinelanguage{maple}{
  keywords={do,end,for,from,local,proc,return,to},
  emph={add,diff,Matrix,MatrixInverse,simplify,taylor},
  comment=[l]{\#}
}
\lstdefinestyle{maplecodestyle}{
  language=maple,
  aboveskip=2mm,
  belowskip=2mm,
  frame=tb,
  breaklines=ture,
  breakindent=0.02\textwidth,
  numbers=left,
  numberstyle=\tiny\color{gray},
  basicstyle=\ttfamily\small,
  keywordstyle=\color{blue}\bfseries\small,
  emphstyle=\color{emphcolor},
  commentstyle=\color{red}
}
\lstnewenvironment{maplecode}[0]{\lstset{style=maplecodestyle}}{}
\lstdefinestyle{mapleinlinestyle}{
  language=maple,
  aboveskip=2mm,
  belowskip=2mm,
  frame=tb,
  breaklines=ture,
  breakindent=0.02\textwidth,
  numbers=left,
  numberstyle=\tiny\color{gray},
  basicstyle=\ttfamily\small,
  keywordstyle=\color{blue}\bfseries\small,
  emphstyle=\color{emphcolor},
  commentstyle=\color{red}
}
\def\mapleinline{\lstinline[style=mapleinlinestyle]}
%%%%%%%%%%%%%%%%%%%%%%%%%%%%%%%%%%%%%%%%%%%%%%%%%%%%%%%%%%%%%%%%%%%%%
\maketitle
\onecolumngrid
\begin{center}
  \textbf{\large 4G: Pure fourth-order gravity}\\
  \vspace{0.05in}
  {\it\large Supplementary Material}\\
  \vspace{0.05in}
  {Shuxun Tian}
\end{center}
\onecolumngrid
\setcounter{equation}{0}
\setcounter{figure}{0}
\setcounter{table}{0}
\setcounter{section}{0}
\setcounter{page}{1}
\makeatletter
\renewcommand{\theequation}{\arabic{equation}}
\renewcommand{\thefigure}{\arabic{figure}}
\renewcommand{\thetable}{\arabic{table}}
\renewcommand{\thesection}{\Roman{section}}
\renewcommand{\thepage}{S\arabic{page}}

This document provides the \href{https://www.maplesoft.com/}{\tt Maple} (version: 2015.0) program that used to simplify Eq. (\ref{eq:29}). The program for Eq. (\ref{eq:05}) can be easily obtained by modifying the existing examples. Section \ref{sec:s01} is dedicated to calculating Eq. (\ref{eq:30}) and Eq. (\ref{eq:31}). Section \ref{sec:s02} shows how to modify the code to calculate Eq. (\ref{eq:32}).

\section{Cosmology}\label{sec:s01}
\leftline{\textbf{Code:}}
\begin{maplecode}
F:=proc(muout,nuout)
  # return the muout-nuout component of Eq. (29) in the main text
  # in comments, $$ means the LaTeX grammar
  local c,a0,t0,alpha,beta,lambda,mu,nu,rho,coords,g,ginv,gpartial,Chsym,Rie,Ric,R,DR,DDR,BoxR,DRic,DDRic,BoxRic,H4,H5,Huv,Fuv,HminusF,r1,r2,r3;
  coords:=[x,y,z,t];
  # g[mu,nu]=$g_{\mu\nu}$
  g:=Matrix(4,4,0);
  g[1,1]:=(a0*(t/t0)^n)^2;
  g[2,2]:=g[1,1];
  g[3,3]:=g[1,1];
  g[4,4]:=-c^2;
  # ginv[mu,nu]=$g^{\mu\nu}$
  ginv:=MatrixInverse(g);
  # gpartial[mu,nu,alpha]=$\partial_\alpha g_{\mu\nu}$
  for mu from 1 to 4 do
    for nu from 1 to 4 do
      for alpha from 1 to 4 do
        gpartial[mu,nu,alpha]:=diff(g[mu,nu],coords[alpha]);
  end do;end do;end do;
  # Chsym[lambda,mu,nu]=$\Gamma^\lambda_{\mu\nu}$
  for lambda from 1 to 4 do
    for mu from 1 to 4 do
      for nu from 1 to 4 do
        Chsym[lambda,mu,nu]:=1/2*add(ginv[lambda,alpha]*(gpartial[mu,alpha,nu]+gpartial[nu,alpha,mu]-gpartial[mu,nu,alpha]),alpha=1..4);
  end do;end do;end do;
  # Rie[rho,lambda,mu,nu]=$R^\rho_{\ \lambda\mu\nu}$
  for rho from 1 to 4 do
    for lambda from 1 to 4 do
      for mu from 1 to 4 do
        for nu from 1 to 4 do
          r1:=add(Chsym[rho,alpha,mu]*Chsym[alpha,lambda,nu],alpha=1..4);
          r2:=add(Chsym[rho,alpha,nu]*Chsym[alpha,lambda,mu],alpha=1..4);
          Rie[rho,lambda,mu,nu]:=diff(Chsym[rho,lambda,nu],coords[mu])-diff(Chsym[rho,lambda,mu],coords[nu])+r1-r2;
  end do;end do;end do;end do;
  # Ric[mu,nu]=$R_{\mu\nu}$
  # R is the Ricci scalar
  R:=0;
  for mu from 1 to 4 do
    for nu from 1 to 4 do
      Ric[mu,nu]:=add(Rie[alpha,mu,alpha,nu],alpha=1..4);
      R:=R+ginv[mu,nu]*Ric[mu,nu];
  end do;end do;
  # DR[nu]=$\nabla_\nu R$
  for nu from 1 to 4 do
    DR[nu]:=diff(R,coords[nu]);
  end do;
  # DDR[mu,nu]=$\nabla_\mu\nabla_\nu R$
  # BoxR=$\Box R$
  BoxR:=0;
  for mu from 1 to 4 do
    for nu from 1 to 4 do
      r1:=add(Chsym[lambda,mu,nu]*DR[lambda],lambda=1..4);
      DDR[mu,nu]:=diff(DR[nu],coords[mu])-r1;
      BoxR:=BoxR+ginv[mu,nu]*DDR[mu,nu];
  end do;end do;
  # H4[mu,nu]=$R_{\alpha\mu\nu\beta}R^{\alpha\beta}$
  for mu from 1 to 4 do
    for nu from 1 to 4 do
      H4[mu,nu]:=0;
      for rho from 1 to 4 do
        for lambda from 1 to 4 do
          for beta from 1 to 4 do
            H4[mu,nu]:=H4[mu,nu]+ginv[lambda,beta]*Rie[rho,mu,nu,beta]*Ric[rho,lambda];
  end do;end do;end do;end do;end do;
  # H5=$R_{\alpha\beta}R^{\alpha\beta}$
  H5:=0;
  for rho from 1 to 4 do
    for lambda from 1 to 4 do
      for alpha from 1 to 4 do
        for beta from 1 to 4 do
          H5:=H5+ginv[rho,alpha]*ginv[lambda,beta]*Ric[alpha,beta]*Ric[rho,lambda];
  end do;end do;end do;end do;
  # DRic[mu,nu,beta]=$\nabla_\beta R_{\mu\nu}$
  for mu from 1 to 4 do
    for nu from 1 to 4 do
      for beta from 1 to 4 do
        r1:=add(Chsym[lambda,beta,mu]*Ric[lambda,nu],lambda=1..4);
        r2:=add(Chsym[lambda,beta,nu]*Ric[mu,lambda],lambda=1..4);
        DRic[mu,nu,beta]:=diff(Ric[mu,nu],coords[beta])-r1-r2;
  end do;end do;end do;
  # DDRic[mu,nu,alpha,beta]=$\nabla_\alpha\nabla_\beta R_{\mu\nu}$
  # BoxRic[mu,nu]=$\Box R_{\mu\nu}$
  for mu from 1 to 4 do
    for nu from 1 to 4 do
      BoxRic[mu,nu]:=0;
      for alpha from 1 to 4 do
        for beta from 1 to 4 do
          r1:=add(Chsym[lambda,alpha,mu]*DRic[lambda,nu,beta],lambda=1..4);
          r2:=add(Chsym[lambda,alpha,nu]*DRic[mu,lambda,beta],lambda=1..4);
          r3:=add(Chsym[lambda,alpha,beta]*DRic[mu,nu,lambda],lambda=1..4);
          DDRic[mu,nu,alpha,beta]:=diff(DRic[mu,nu,beta],coords[alpha])-r1-r2-r3;
          BoxRic[mu,nu]:=BoxRic[mu,nu]+ginv[alpha,beta]*DDRic[mu,nu,alpha,beta];
  end do;end do;end do;end do;
  # muout-nuout component of the field equations
  # Huv for Eq. (5); Fuv for Eq. (3); HminusF for Eq. (29)
  Huv:=BoxRic[muout,nuout]+1/2*g[muout,nuout]*BoxR-DDR[muout,nuout]-2*H4[muout,nuout]-1/2*g[muout,nuout]*H5;
  Fuv:=R*Ric[muout,nuout]-g[muout,nuout]*R^2/4-DDR[muout,nuout]+g[muout,nuout]*BoxR;
  HminusF:=BoxRic[muout,nuout]-1/2*g[muout,nuout]*BoxR+1/4*g[muout,nuout]*R^2-1/2*g[muout,nuout]*H5-2*H4[muout,nuout]-R*Ric[muout,nuout];
  return(simplify(HminusF));
end proc:
\end{maplecode}%

\leftline{\textbf{Examples:}}
\newcommand\inputsymbol{{\fontsize{7pt}{0pt}\selectfont$>\ \ $}}
\noindent\inputsymbol\mapleinline{with(LinearAlgebra):}\\
\noindent\inputsymbol\mapleinline{Code. # paste the Code here, and press Enter}\\
\noindent\inputsymbol\mapleinline{F(4,4) # 00 component}
{\color{blue}
\begin{equation}
  \frac{9n^2(2n-1)}{c^2t^4}
\end{equation}}%
\noindent\inputsymbol\mapleinline{F(1,1) # 11 component}
{\color{blue}
\begin{equation}
  -\frac{3a_0^2n(6n^2-11n+4)}{c^4t^4}\cdot\left(\frac{t}{t_0}\right)^{2n}
\end{equation}}%

\section{Newtonian approximation}\label{sec:s02}
\renewcommand\linenumberfont{\tiny\color{gray}}
\setlength\linenumbersep{10pt}
\setlength\parindent{1em}
\leftline{\textbf{Code:}}
\vspace{0.21cm}
\nolinenumbers\hrule\par
\vspace{0.16cm}
\linenumbers[1]\noindent\mapleinline{F:=proc(muout,nuout)}\par
\nolinenumbers$\cdots\ $\mapleinline{# ellipsis indicates the code in these lines remains unchanged}\par
\linenumbers[8]\mapleinline{g[1,1]:=1-2*epsilon*Phi(x,y,z)/c^2; # epsilon is a math infinitesimal used to do Taylor expansion}\par
\nolinenumbers$\cdots$\par
\linenumbers[11]\mapleinline{g[4,4]:=-c^2(1+2*epsilon*Phi(x,y,z)/c^2);}\par
\nolinenumbers$\cdots$\par
\linenumbers[99]\mapleinline{return(simplify(taylor(HminusF,epsilon,2)));}\par
\linenumbers[100]\noindent\mapleinline{end proc:}\par
\vspace{0.16cm}
\nolinenumbers\hrule\par
\vspace{0.4cm}

\leftline{\textbf{Examples:}}
\noindent\inputsymbol\mapleinline{with(LinearAlgebra):}\\
\noindent\inputsymbol\mapleinline{Code. # paste the Code here, and press Enter}\\
\noindent\inputsymbol\mapleinline{F(4,4) # 00 component}
{\color{blue}
\begin{equation}
  \left(2\frac{\p^4\Phi}{\p x^4}+2\frac{\p^4\Phi}{\p y^4}+2\frac{\p^4\Phi}{\p z^4}+4\frac{\p^4\Phi}{\p x^2\p y^2}+4\frac{\p^4\Phi}{\p x^2\p z^2}+4\frac{\p^4\Phi}{\p y^2\p z^2}\right)\varepsilon+O(\varepsilon^2)
\end{equation}}%
\noindent\inputsymbol\mapleinline{F(1,1) # 11 component}
{\color{blue}
\begin{equation}
  O(\varepsilon^2)
\end{equation}}%

\end{document}